\documentclass[grl]{agutex}

 \usepackage{lineno}

\usepackage[dvips]{graphicx}

\authorrunninghead{Garc\'ia Mu\~noz et al.}

\titlerunninghead{Global imaging of the Venus O$_2$ visible nightglow}

\begin{document}

\title{Limb
imaging of the Venus O$_2$ visible nightglow 
with the Venus Monitoring Camera
}

\authors{
A. Garc\'ia Mu\~noz,\altaffilmark{1,2,3}
R. Hueso,\altaffilmark{4,5} 
A. S\'anchez-Lavega,\altaffilmark{4,5}
W. J. Markiewicz,\altaffilmark{6}
D. V. Titov,\altaffilmark{1}
O. Witasse,\altaffilmark{1}
and
A. Opitz\altaffilmark{1,2}
}

 \altaffiltext{1}{ESA/RSSD, ESTEC, 2201 AZ Noordwijk, The Netherlands} 
 \altaffiltext{2}{ESA Fellow} 
 \altaffiltext{3}{Grupo de Ciencias Planetarias, Dpto. de F\'isica Aplicada I,
 ETS Ingenier\'ia, UPV-EHU, Alameda Urquijo s/n, 48013 Bilbao, Spain} 

 \altaffiltext{4}{Dpto. F\'isica Aplicada I, ETS Ingenier\'ia, UPV-EHU, Alameda Urquijo s/n, 48013 
 Bilbao, Spain}
 \altaffiltext{5}{Unidad Asociada Grupo Ciencias Planetarias UPV/EHU-IAA (CSIC)}
 \altaffiltext{6}{
 Max Planck Institute for Solar System Research, Max-Planck-Str. 2, 37191 Katlenburg-Lindau, Germany}

%\newpage 
\begin{abstract} 

We investigated the Venus O$_2$ visible nightglow with imagery 
from the Venus Monitoring Camera on Venus Express. 
Drawing from data collected between April 2007 and January 2011, 
we study the global distribution of this emission, discovered in the late 70s
by the Venera 9 and 10 missions.
The inferred limb-viewing intensities are on the order of 150 kiloRayleighs at the lower latitudes 
and seem to drop somewhat towards the poles. 
The emission is generally stable, 
although there are episodes when the intensities rise up to 500 kR.
We compare a set of Venus Monitoring Camera observations with 
coincident measurements of the O$_2$ nightglow at 1.27 $\mu$m made 
with the Visible and Infrared Thermal Imaging Spectrometer, also on Venus Express. 
From the evidence gathered in this and past works,  
we suggest a direct correlation between the instantaneous emissions from the 
two O$_2$ nightglow systems. 
Possible implications regarding the uncertain origin of the atomic oxygen 
green line at 557.7 nm are noted.

\end{abstract}

%% ------------------------------------------------------------------------ %%
%
%  BEGIN ARTICLE
%
%% ------------------------------------------------------------------------ %%

% The body of the article must start with a \begin{article} command
%
% \end{article} must follow the references section, before the figures
%  and tables.

\begin{article}

%% ------------------------------------------------------------------------ %%
%
%  TEXT
%
%% ------------------------------------------------------------------------ %%

%\linenumbers*[1]
\section{Introduction}

The Venus atmosphere is a complex chemical system in permanent disequilibrium. 
Some of the occurring reactions are exothermic and lead to
excited-state products that radiate in the form of nightglow.
The list of gases contributing to the Venus nightglow includes 
O, O$_2$, NO and OH, which produce emission features
throughout the ultraviolet, visible and near-infrared spectrum 
\citep{connesetal1979,feldmanetal1979,garciamunozetal2009a,krasnopolskyetal1977, 
piccionietal2008,slangeretal2001}. 

The Herzberg II system of O$_2$ is one of such nightglow emissions, 
resulting from the radiative transition 
O$_2$ $c$(0)$\rightarrow$$X$($v$'')+$h\nu$ between 350 and 700 nm. 
The emission was first detected at Venus
by the spectrometers on the Venera 9 and 10 missions
\citep{krasnopolskyetal1977} and assigned to the Herzberg II system soon after
\citep{lawrenceetal1977}.
Typically, the emitting layer is $\sim$15--20 km thick and centered at 
90--100 km altitude. Mean values for the overhead intensity are in the range of
2--3 kiloRayleighs (or about 50 times more in limb viewing through the layer
center). 
The Venera spectra did also contain the signature of the O$_2$ Chamberlain system,  
$A'$(0)$\rightarrow$$a$($v''$)+$h\nu$, 
the ratio of $c$--$X$ to $A'$--$a$ intensities being about 15 to 1 \citep{krasnopolsky1983}. 
We refer to the joint emission from the $c$--$X$ to $A'$--$a$
systems as the Venus O$_2$ visible nightglow.

There is compelling evidence that the O$_2$ nightglow \citep{slangercopeland2003}, 
including the $a$(0)--$X$($v''$= 0, 1) bands at 1.27 and 1.58 $\mu$m, respectively, 
arises from one or more O$_2^*$ states excited in the 
three-body reaction:
\begin{equation}
\rm{O} + \rm{O} + \rm{M} \rightarrow \rm{O}_2^* + \rm{M}.
\label{termolecular_eq}
\end{equation}
The oxygen atoms are produced by photodissociation of CO$_2$ on the planet's dayside and 
transported to the nightside by subsolar-to-antisolar winds in the upper atmosphere
\citep{foxbougher1991,bougheretal2006}. 
In spite of decades of laboratory and planetary (including Earth and Mars)
investigations, the understanding of the 
mechanisms that channel the O$_2^*$ product(s) from the reaction of Eq. (\ref{termolecular_eq})
into the $c$(0), $A'$(0) and $a$(0) emitting states remains incomplete. 

Indeed, it is remarkable that in the Venus atmosphere the 
strong O$_2$ $a$(0)--$X$(0) emission at 1.27 $\mu$m (mean overhead intensity 
of about 1 MR at equatorial latitudes, dropping to  $\sim$0.2 MR at 
60--70$^{\circ}$N, as reported by \citet{piccionietal2009} and supported in related
work
\citep{drossartetal2007,baileyetal2008,gerardetal2009,soretetal2012}) 
varies locally by more than an order of magnitude on time scales of hours 
(especially in nadir viewing)
\citep{crispetal1996,huesoetal2008,piccionietal2009}.
However, reported intensities for the O$_2$ visible nightglow (mostly in limb
viewing) fall typically within a range of 2--3 from 150 kR
 \citep{krasnopolskyetal1977,
bougherborucki1994,slangeretal2001,slangeretal2006,slangeretal2012, gerardetal2013, migliorinietal2013}. 

In the current work, we take a new look at the Venus O$_2$ visible nightglow 
by means of imagery of the planet's limb obtained with
the Venus Monitoring Camera (VMC) on Venus Express (VEx). 
The experiment is unique in that it provides snapshot observations over
a baseline of nearly four years, 
longer than past space-based efforts (namely, the Venera 9 and 10
missions, and Pioneer Venus Orbiter, PVO). 
Unlike PVO, that used the on-board star tracker  
for monitoring the nightglow \citep{bougherborucki1994}, 
VMC provides vertical resolution of the emitting layer. 
The emerging picture of the Venus O$_2$ visible nightglow resembles
that of the O$_2$ nightglow at 1.27 $\mu$m. 
A few coincident observations of VMC and the
Visible and Infrared Thermal Imaging Spectrometer
(VIRTIS, also on VEx) at 1.27 $\mu$m support the conclusion. 
VMC sets a valuable reference for imaging the nightglow on 
planets other than Earth.

\section{Selection and treatment of images}

%VEx was launched in November 2005, and set into an elliptical, nearly-polar, 24-hour 
%orbit  around Venus in May 2006 \citep{svedhemetal2007}, 
%having since circled the planet over 2,000 times.
%In an orbit, 
%the planet-to-spacecraft distance ranges from about 165--300 km above the planet's surface 
%at periapsis to 11 planetary radii at apoapsis, which allows for a variety
%of scientific objectives to be addressed.

VMC is the wide-angle camera on VEx  \citep{markiewiczetal2007}. 
The VMC optical system projects a
$\sim$17.5$^{\circ}$-field of view through four separate spectral 
channels onto the four quadrants of a CCD detector. 
The angular resolution is about 0.7 mrad/pixel, which translates into
2.3-5.5 km/pixel at the tangent point 
for limb viewing and spacecraft planetocentric distances of 7,000--10,000 km.
The visible filter is sensitive in the 500--560 nm region, which means it is well
suited for detection of the O$_2$ $c$(0)--$X$($v''$=9, 10) bands.

%For our investigation, % of the Venus O$_2$ visible nightglow with VMC, 
We considered the full public archive of VMC images in the visible  
channel,\footnote{ftp://psa.esac.esa.int/pub/mirror/VENUS-EXPRESS/VMC/}  
which goes from orbits 24 (15/05/2006) to 2099 (18/01/2012). 
Visual examination of the images including the planet's limb show a faint though 
distinct nightglow layer near 100 km altitude when the longest exposure times allowed
for by the instrument (30 secs) are utilized. From each image, we obtained 5--6 bins, each 
bin representing a local time/latitude region on the image,  
and the corresponding vertical emission profiles. 
For details on the navigation with the PLIA software \citep{huesoetal2010} 
and radiometric calibration, see the 
Supplementary Material.
The VMC intensities quoted here refer to the full O$_2$ $c$(0)--$X$($v''$) 
progression in limb viewing.

We produced a total of 515 bins from 117 images obtained 
over 38 orbits.
The first and last orbits analyzed correspond to numbers 317 and 1716 in the
VEx numbering scheme, and date from 04/03/2007 and 01/01/2011, respectively. 
Solar activity in that period is low, with solar radio fluxes  $F_{10.7\rm{cm}}$=70--90. 
There are more images probing the northern hemisphere 
($\sim$2/3 of the total), but overall both hemispheres are comparably well sampled. 
Often, the images form sequences
that probe significant areas of the planet's nightside in a single orbit. 
Fig. (I, Top) in the Supplementary Material 
displays six of those images, exploring 
the southern hemisphere before periapsis in orbit 1027, whereas
Fig. (I, Bottom) shows the range of local times and latitudes probed
over the sequence.

Figure (II) in the Supplementary Material shows (black symbols) 
the local time and latitude at the tangent point for all bins. 
Latitudes range from 60$^{\circ}$S to 65$^{\circ}$N  
and local times are largely (but not exclusively) within two hours 
from midnight. 
The red symbols are specific to instances of high nightglow emission. 
The green and blue symbols represent coincident VMC and VIRTIS 
observations, respectively. These cases are discussed below. 

\section{Results}

\begin{figure}[h]
\noindent\includegraphics[width=20pc]{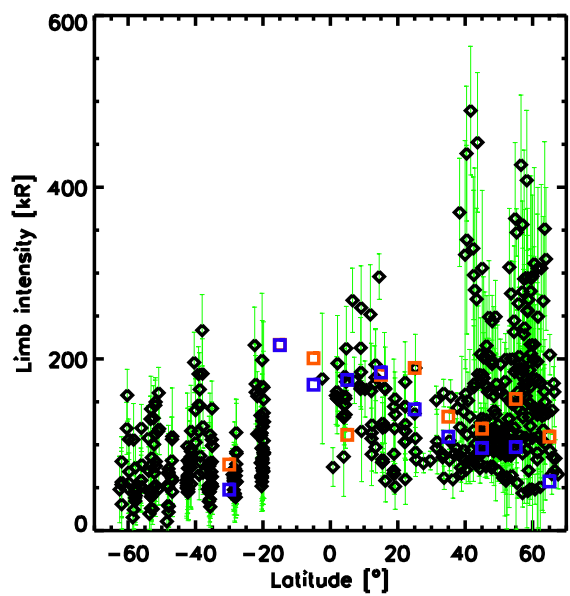}
\caption{
Black diamonds: Limb-integrated intensity against latitude for the full set of
analyzed VMC bins.
The uncertainty bars are standard deviations in the intensity at the altitude of the emission peak
(see text and Fig. (\ref{profile_fig})).
Blue squares: VIRTIS limb-integrated intensity (scaled by 1/200) 
for the O$_2$ $a$(0)--$X$(0) band at
1.27 $\mu$m as published in \citet{garciamunozetal2009b}. Red squares: VIRTIS limb-integrated
intensity for the O$_2$ visible nightglow. 
\label{ivslati_fig}
}
\end{figure}

Figure (\ref{ivslati_fig}) shows VMC peak intensities against latitude 
at the tangent point.
The mean limb-viewing intensity in the 20$^{\circ}$S--20$^{\circ}$N range is $\sim$150 kR, 
which is reasonably consistent with the intensities found by the Veneras  
\citep[$\sim$(1.5--3)$\times$50=$\sim$(75--150) kR, ][]{krasnopolskyetal1977,krasnopolsky1983}, 
PVO  \citep[$\sim$(1.5--5.7)$\times$50=$\sim$(75--285) kR, ][]{bougherborucki1994}, 
VEx/VIRTIS \citep[$\sim$(100--200) kR, ][]{garciamunozetal2009b,gerardetal2013,migliorinietal2013} and 
ground-based observations \citep[$\sim$(3--5)$\times$50=$\sim$(150--250) kR, ][]{slangeretal2001,slangeretal2006}.
For reference, the figure includes the average intensities from 90 to 100 km altitude
for both the visible and near-infrared emissions as measured with VIRTIS 
between January 2007 and May 2008 \citep{garciamunozetal2009b}. 
The latter are scaled 
by an estimated ratio for the visible and near-infrared emissions of 
1/200. 
The VIRTIS data conform to the canonical view that the O$_2$ nightglow peaks near the
antisolar point because the recombining oxygen atoms are transported from the dayside
in the subsolar-to-antisolar circulation pattern \citep{bougheretal2006, brechtetal2011}.

Our interpretation of Fig. (\ref{ivslati_fig}) is that the VMC intensities generally decay 
with latitude, a view that follows that for the averaged VIRTIS data. 
The conclusion is more easily justified for the southern hemisphere, 
but is also consistent with a large number of the VMC measurements over the northern hemisphere. 
Thus, the dispersion in the VMC intensities, particularly north of 40$^{\circ}$N,  
likely reflects temporal and spatial variations, such as those known to occur at 1.27 $\mu$m.
A few brighter-than-average episodes are found in orbits 352, 364 and 1286
(see Fig. II, red symbols), having peak intensities that reach above 350 kR, 
which has an impact on the trend of VMC points in Fig. (\ref{ivslati_fig}). 

The lack of symmetry between the two hemispheres, even in the instances of 
moderate emission, may be caused by the different vertical resolution in the measurements
north and south of the equator. The VEx orbit imposes that the VMC spatial resolution
at the limb improves as the spacecraft nears the northern polar latitudes, 
which leads to a progressive positive bias in the corresponding 
peak intensities.
 
\begin{figure}[ht]
 \noindent\includegraphics[width=20pc]{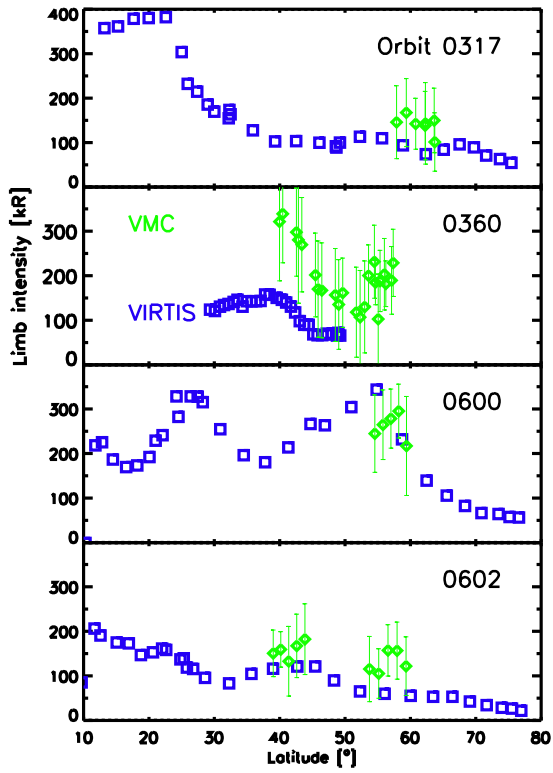}
\caption{Limb-integrated intensity against latitude for orbits 317, 360, 600 and
602. Green diamonds: VMC observations, and associated uncertainties. Blue squares: Coincident VIRTIS
observations of the O$_2$ $a$(0)--$X$(0) band at 1.27 $\mu$m (scaled by 1/200). 
Uncertainties in the VIRTIS measurements are less than for VMC, and therefore omitted.
\label{ipanel_fig} 
}
\end{figure}

The optical axes of both the VMC and VIRTIS instruments are aligned along the +z-axis
of VEx, which makes it possible to obtain simultaneous, co-located observations of 
the Venus atmosphere. 
We identified four orbits (317, 360, 600 and 602) for which we can directly 
compare the O$_2$ visible nightglow, as observed with VMC, and the 
O$_2$ near-infrared nightglow, as observed with VIRTIS. 
Table (\ref{VIRTIS_table}) summarizes some relevant information 
about the VMC bins in those orbits.
Figure (\ref{ipanel_fig}) shows the two sets of intensities against latitude, with green and blue 
symbols representing VMC and VIRTIS data, respectively. 
Local times and latitudes specific to the VMC and VIRTIS 
observations in the four orbits are shown in Fig. (II).
Both sets of intensities appear reasonably consistent, 
especially where the two instruments probe overlapping regions 
(i.e., orbits 317, 600, 602). 
Orbit 602 exhibits moderate enhancements in the two emissions, 
when the VMC intensities reach 250--300 kR. 
The comparison would obviously benefit from a longer record of simultaneous measurements, 
especially if conditions of high emission could be identified.

\begin{figure}[h]
\noindent\includegraphics[width=20pc]{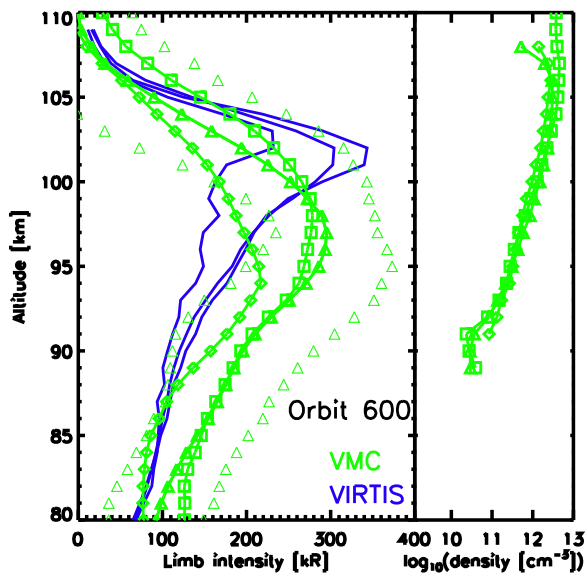}
\caption{Left. Limb-integrated intensity against tangent altitude for three
bins from orbit 600 (thick triangles, squares and diamonds), where intensities are relatively high. 
Each bin corresponds to a slightly different latitude, as seen in 
Fig. (\ref{ipanel_fig}). 
Both VMC profiles of the O$_2$ visible nightglow and 
VIRTIS profiles of the O$_2$ $a$(0)--$X$(0) band at 1.27 $\mu$m 
(scaled by 1/200) are shown. Uncertainties at each altitude are given for the VMC 
profile with triangle symbols and shown with disconnected triangles. For any one bin, 
the uncertainties given in Figs. (\ref{ivslati_fig}) and (\ref{ipanel_fig}) are
the uncertainties at the altitude of peak emission of the corresponding profile.
The mismatch in the altitude of the visible and near-infrared 
emission peaks is likely affected by the different
integration times and spatial resolutions of the two instruments. 
Right. Atomic oxygen profiles retrieved from the VMC profiles on the left panel.
\label{profile_fig}
}
\end{figure}

Figure (\ref{profile_fig}, Left) shows the emission vertical profiles for VMC and VIRTIS data in
a few selected examples from orbit 600.
The VMC profiles are visibly broader than the VIRTIS ones, which likely results 
from three factors specific to VMC, namely 
its poorer spatial resolution (the VIRTIS instantaneous field of view is 2.5$\times$10$^{-4}$ rad/pixel), 
the co-addition of vertical profiles obtained with relatively long
exposure times, 
and a point spread function of a few pixels \citep{titovetal2012}. 
It is difficult to estimate the accuracy in the absolute 
vertical scale of the
VMC profiles and, thus, we have focused on the magnitudes of the emission peaks.
Uncertainty bars for the intensities are given for one VMC profile. 
They are calculated, at each altitude, 
as the standard deviation of the intensity from profiles that 
contribute to the bin. 
The uncertainty bars are affected by the three factors mentioned
above, but they also provide a measure of
the photon statistics and the spatial variability 
of the nightglow emission within the bin. 
The intensities and uncertainty bars shown in Figs. (\ref{ivslati_fig})--(\ref{ipanel_fig}) 
are specific to the altitude of peak emission.

Nightglow observations may be used for remotely sensing the atmospheric atomic oxygen. 
As a demonstration, we retrieved the atom number densities for the three VMC emission profiles 
in Fig. (\ref{profile_fig}, Left)
with the energy-transfer parameters for excitation and quenching of the 
Venus O$_2$ visible nightglow derived by \citet{garciamunozetal2009b}. 
The profiles are plotted in Fig. (\ref{profile_fig}, Right), and show atomic oxygen 
number densities of 10$^{11}$--10$^{12}$  cm$^{-3}$ at 100 km. 
Such values are somewhat larger than, but in reasonable agreement with, the 
retrievals from the O$_2$ nightglow at 1.27 $\mu$m \citep{garciamunozetal2009b, gerardetal2009}
at the same altitude. The broad emission profiles lead to somewhat 
overestimated atomic oxygen number densities at the higher altitudes.

%\newpage
\section{Discussion}

%The inferred intensities generally conform with those reported from previous in-orbit
%investigations and ground-based observations.
%The capacity of VMC for simultaneous temporal, latitudinal/longitudinal 
%and vertical resolution 
%means a valuable complement and expansion to past efforts. Thus, 
%VMC sets a valid reference for O$_2$ nightglow imagers on future Venus missions. 
The VMC observations analyzed here show that the O$_2$ visible nightglow is
a permanent feature of the Venus nightside. The intensities seem to
peak at low latitudes and drop polewards, which is consistent with past 
O$_2$ visible and near-infrared nightglow observations with VIRTIS \citep{garciamunozetal2009b}.  
The nightglow is variable, as demonstrated by occasional high emission events. 

The main conclusion is that the 
O$_2$ visible nightglow resembles the O$_2$ nightglow at 1.27 $\mu$m. 
This suggests, in turn, a direct correlation between the two, which is not 
unexpected since both emissions can ultimately be traced to the recombination reaction of 
Eq. (\ref{termolecular_eq}).
In this respect, we note the importance of the viewing geometry 
when comparing various measurements. 
Most pre-VEx observations of the
O$_2$ nightglow at 1.27 $\mu$m came from nadir observations \citep[e.g.]{baileyetal2008,
connesetal1979,crispetal1996}. However, pre-VEx space-based 
observations of the O$_2$ visible nightglow were typically conducted in limb viewing
\citep{krasnopolskyetal1977,bougherborucki1994}, 
which tends to smooth out localized features and therefore minimize changes. 

The tentative correlation may bear implications for the excitation of the 
atomic oxygen green line at 557.7 nm \citep{slangeretal2001}. 
By now, the green line has been detected several
times, always from the ground \citep{slangeretal2006,slangeretal2012,grayetal2012}, 
its intensity being permanently more than tenfold fainter 
than the 170 R of overhead intensity on the occasion of its discovery.
%It took decades to establish that the terrestrial green line 
%is excited in the transfer of 4.2 eV from an O$_2$ precursor state to the oxygen atom
%by the so-called Barth mechanism \citep{barthhildebrandt1961, steadmanthrush1994}. 
%An important limitation of this mechanism in the Venus atmosphere is that the
%green line seems more variable than the population of the $c$(0) or $A'$(0) states and,
%probably, than the population of unobserved O$_2$ states with energies above 4.2 eV. 
We cannot prove a connection between the O$_2$ visible nightglow and the green line 
(such a connection would hint at a mesospheric origin of the latter),  
and cannot either explain why the overhead intensity for the O$_2$ $c$--$X$ system was 
relatively normal (5.1 kR, i.e. about 250 kR in limb viewing) on the occasion of the 
Venus green line discovery. However, it appears
that the O$_2$ visible nightglow may be as variable as the associated emission at 1.27 $\mu$m.
A usual objection for the green line originating in the mesosphere 
in the interaction with an energetic O$_2$ precursor state is that the 
relatively energetic O$_2$ $c$(0) state was considered 
to undergo little variability.
The realisation that the O$_2$ visible nightglow may be as variable as the
near-infrared emission 
eliminates a fundamental impediment to a mesospheric origin of the atomic 
oxygen green line.  
The realisation, however, is not sufficient to prove a direct connection between the O$_2$ and O nightglow emissions 
or to rule out the other mesospheric and ionospheric mechanisms for excitation of the green line
proposed in recent years \citep{slangeretal2006,slangeretal2012,fox2012}.

%The O$_2$ visible nightglow may serve as a tracer of atmospheric winds by tracking
%localized emission features as they evolve, 
%in the way it is done in the near infrared \citep{bougherborucki1994,huesoetal2008}. 
%The advantages of the visible emission 
%include the short radiative lifetime of the O$_2$ $c$(0) state, 3.5 secs (much less than about 
%one hour for the $a$(0) state), and the absence of thermal radiation at visible wavelengths 
%blending with the nightglow emission.
%Additional wind measurements are required for understanding the major circulation 
%patterns and the role of gravity waves in the Venus atmosphere \citep{bougheretal2006}, 
%which could be addressed by a future O$_2$ visible nightglow imager.

%Finally, we note that further investigation of the Venus O$_2$ visible
%nightglow might be possible in the near future either with the VEx star trackers \citep{svedhemkoschny2012} or
%with the Lightning and Airglow Camera \citep{takahashietal2008}
%when the Akatsuki spacecraft is inserted into orbit after a
%failed attempt in 2010.

\begin{acknowledgments}
RH and ASL were supported by the Spanish MICINN project AYA2009-10701 and
AYA2012-36666 with FEDER support, Grupos Gobierno Vasco IT-464-07 and UPV/EHU UFI11/55.
AGM gratefully acknowledges the assistance of Teresa del R\'io-Gaztelurrutia (UPV/EHU, 
Bilbao, Spain) with PLIA.
\end{acknowledgments}

\end{article}

%\cleardoublepage

%% Enter Figures and Tables here:

% When submitting articles through the GEMS system:
% COMMENT OUT ANY COMMANDS THAT INCLUDE GRAPHICS.

\begin{table} 
\caption{\label{VIRTIS_table} Image number, local time (LT), latitude (Lat), peak 
intensity (I$_{\rm{peak}}$) and altitude at the intensity peak (z$_{\rm{peak}}$) for selected 
VMC bins in orbits 317, 600 and 602, for which there are simultaneous VIRTIS observations.
 }
\begin{flushleft}
\begin{tabular}{ccccc}
\hline
\hline
\multicolumn{1}{c}{Image} & LT & Lat & I$_{\rm{peak}}$ [kR] & z$_{\rm{peak}}$ [km] \\
%$r_b-R_{\venus}$ & $\nu_{\lambda=0.55\mu m}$ & $H_{\rm{atm}}$  & $\alpha_{\rm{refr}}$ & $\alpha_{\rm{refr}}^*$ \\
%\multicolumn{1}{c}{[km]} & \multicolumn{1}{c}{} & [km] &\multicolumn{2}{c}{[arc sec]} \\
\hline
  317$\_$89 & 24.6 & 63.7 & 101.3 &  93 \\
%  317$\_$89 & 24.5 & 62.3 & 138.3 & 95 \\
  317$\_$89 & 24.5 & 60.8 & 142.3 &  95 \\
%  317$\_$89 & 24.4 & 59.4 & 167.5 & 97 \\
  317$\_$89 & 24.3 & 57.9 & 145.9 & 100 \\
  317$\_$90 & 24.6 & 63.6 & 149.8 &  97 \\
  317$\_$90 & 24.5 & 62.3 & 143.2 &  99 \\
\hline
  600$\_$98 & 23.1 & 59.4 & 217.0 &  94 \\
%  600$\_$98 & 23.1 & 58.2 & 295.3 & 97 \\
  600$\_$98 & 23.2 & 57.0 & 278.4 &  98 \\
%  600$\_$98 & 23.2 & 55.8 & 264.7 & 97 \\
  600$\_$98 & 23.2 & 54.6 & 244.7 &  97 \\
\hline  
  602$\_$175 & 23.5 & 43.9 & 182.4 & 101 \\
%  602$\_$175 & 23.5 & 42.6 & 167.4 & 99 \\
  602$\_$175 & 23.5 & 41.4 & 132.8 &  98 \\
%  602$\_$175 & 23.5 & 40.2 & 159.4 & 93 \\
  602$\_$175 & 23.5 & 39.0 & 150.7 &  90 \\ 
  602$\_$176 & 23.3 & 59.3 & 121.9 &  93 \\
%  602$\_$176 & 23.4 & 57.9 & 156.7 & 95 \\
  602$\_$176 & 23.4 & 56.6 & 157.3 &  95 \\
%  602$\_$176 & 23.4 & 55.1 & 105.3 & 95 \\
  602$\_$176 & 23.5 & 53.7 & 115.4 &  96 \\
\hline
\hline
%\hline
%\multicolumn{5}{l}{
%$\alpha_{\rm{refr}}^*$= 
%$\nu_{\lambda=0.55\mu m}$$\sqrt{2\pi R_{\venus}/H_{\rm{atm}}}$.  
%} \\
%\tableline
%\tableline
\end{tabular}
\end{flushleft}
\end{table}

%\begin{figure}[h]
% \noindent\includegraphics[width=20pc]{./FIGURES/emitlayer1027.eps}
%\\ 
% \noindent\includegraphics[width=20pc]{./FIGURES/diskfig.eps} 
%\caption{
%\label{layer_fig} Top. Sequence of six VMC images of the southern hemisphere
%from orbit 1027. % (11/02/2009). 
%The nightglow layer stands out when
%looking through the planet's limb near 100 km altitude. 
%From 500 to 560 nm, the planet's thermal emission that reaches the top of 
%the atmosphere is negligible. As a consequence, 
%the on-disk and off-disk portions of the images are nearly
%indistinguishable. To improve the photon statistics, we resampled and co-added
%the signal from contiguous cuts of the nightglow layer into bins of average
%properties. The layer in image 1027$\_$0044,
%which probes as far as 75$^{\circ}$S, is notably brighter than in the other images,
%possibly because there is a contribution from scattered sunlight. That image was 
%omitted from the scientific analysis. Bottom. Range of local times and latitudes 
%at the tangent point for each image.
%}
%\end{figure}

%\begin{figure}[h]
%% \noindent\includegraphics[width=20pc]{./FIGURES/mapping.eps}
%\caption{Diamonds represent 
%local time and latitude at the tangent point for the full set of 
%VMC bins analyzed. Green diamonds identify the subset of VMC bins for which there
%are coincident VIRTIS observations,  
%the latter shown with blue squares. 
%Red diamonds represent the VMC bins with intensities higher than 350 kR. 
%\label{mapping_fig} }
%\end{figure}

\end{document}